\documentstyle[12pt]{article}
\def\singlespace{\def\baselinestretch{1}\def\arraystretch{1}\@normalsize}

\def\gcm2{\,{\rm g\,cm}^{-2}}

\def\la{\mathrel{\mathpalette\fun <}}
\def\ga{\mathrel{\mathpalette\fun >}}
\def\etal {{\it et~al.}}
\def\fun#1#2{\lower3.6pt\vbox{\baselineskip0pt\lineskip.9pt
  \ialign{$\mathsurround=0pt#1\hfil##\hfil$\crcr#2\crcr\sim\crcr}}}
\begin{document}
\pagestyle{empty}
\begin{center}
{\Large \bf Intergalactic Matter And Cocoons 
of Radio Galaxies
}

\vspace{.3in}

Biman B.~Nath$^{1,2}$\\

\vspace{0.2in}
\centerline{$^{1}$Inter-University Center for Astronomy \& Astrophysics,
Post Bag 4, Pune 411007, India $^\ast$}
 \centerline{$^{2}$Max-Planck-Institut f\"ur Radioastronomie,
Auf dem H\"ugel 69, D-53121 Bonn, Germany}
\noindent
$^\ast$ Present address

\vspace {.2in}
\centerline{To appear in Monthly Notices of RAS}

\end{center}

\vspace{0.3in}

\begin{abstract}
The cocoons surrounding powerful radio sources can be extensive
if the jet that feeds the cocoon is light and supersonic. They have
been shown to remain overpressured with respect to the 
ambient medium for most of the life time of the sources. The observed 
lobes of the radio sources form
parts of these extensive cocoons. We show that observations 
of the lobes of giant radio sources allow one to estimate the 
{\it density} of the intergalactic medium (IGM) in which the lobes 
are embedded. We estimate the IGM density to be of the order of a 
few percent of the closure density of the universe. We further 
calculate the radio power of the overpressured 
cocoon as a function of time and the ambient density.
\end{abstract}
\vspace{0.2in}
{\it Subject headings:} galaxies: intergalactic medium, jets; cosmology:
miscellaneous.

\vspace{0.7in}
\pagestyle{plain}
\setcounter{page}{1}
\section{Introduction}
The standard scenario for strong double radio sources involves
a `cocoon' surrounding the core and the jet, and consisting of 
shocked ambient medium and shocked jet material (Scheuer 1974; 
Blandford and Rees 1974). It has long been evident that supersonic,
low-density jets deposit most of their energy in the cocoon, which
acts as a ``wastebasket'' (Scheuer 1974). Later studies and numerical
simulations concentrated on the properties of the jet and the
jet/cocoon interface but not much on the evolution of the extensive
cocoon. It was also the general belief that the cocoons would expand 
quickly to reach an equilibrium with the ambient medium.

Recently, however, Begelman and Cioffi (1989; hereafter BC) argued 
that the cocoons remain overpressured with respect to the ambient 
medium for a long time, and that for many sources, they have not yet
reached a state of equilibrium. They then used this fact
to address the problem of jet confinement by the ram
pressure of the cocoon. Their simple model has been 
numerically verified (Loken \etal 1992; Cioffi and Blondin 1992).
The picture that emerges from these studies is that of a 
overpressured, jet-nourished cocoon, whose length and width
depend on the balance of the ram pressure of the ambient medium with,
respectively, the jet's momentum flux and the cocoon pressure.

We will not be concerned with the problem of confining the jet in
the present work,
but rather the physical conditions in the cocoon and in the
ambient medium. Before BC pointed out the tendency of the cocoons 
to remain overpressured, it was generally assumed that the 
pressure in the lobes of giant galaxies (whose lobes lie in 
the IGM, much beyond the hot corona of the host galaxy) also estimated
the pressure of the IGM, as the time for the cocoons to reach
pressure equilibrium with the IGM was thought to be short. 
That, however, does not seem to be the case (see also Subramanyan
and Saripalli 1993).
Here we ask the question whether observations
of the lobes can say anything about the physical conditions of the IGM.

We find that following the simple model of BC, one can estimate
the {\it density} of the ambient medium and one can therefore estimate 
the baryon density of the IGM from the observations of the giant
radio sources. We also calculate the
radio power of the cocoon and its dependences on the density of
the ambient medium and time.

We discuss the density of 
the ambient medium in $\S 3$ and the
radio power in $\S 4$. Throughout the paper, we use a 
Hubble constant of $H= h_{50} 50$ km s$^{-1}$ Mpc $^{-1}$. 
We also express the relevant densities in terms of the closure 
density of the universe (i.e., $\Omega=1$ means a density of 
$\rho=(3 H^2 / 8 \pi G)$. We begin with a brief recapitulation
of the important aspects of the BC model in the next section.

\section{Overpressured Cocoons}

Consider the extensive cocoons produced by light
and hypersonic jets. The lightness of the jet means that the density
of the fluid material in the jet $\rho _j$ is smaller than that of the
ambient medium $\rho _a$; i.e. $\eta = \rho _j / \rho _a < 1$
(hereafter, the subscripts $j, a, c$ will refer to the jet, the 
ambient medium and the cocoon, respectively). We will not consider 
heavy or subsonic jets. Heavy jets behave like `bullets' and do not 
form cocoons. And since the jet luminosity
is proportional to $M_h ^3$ ($M$ is the Mach number of the head of the jet 
with respect to the sound velocity of the ambient medium), high
luminosity of the classical double jet implies hypersonic,
if not supersonic, jets.

The jet power $L_j$ is assumed to be constant in time, for
simplicity. The velocity of the head of the jet $v_h$ is 
determined by the balance of the thrust of the jet ($\sim L_j / 
v_j$; we assume $\beta _j =v_j/c \sim 1$, as in BC) 
spread over the
cross sectional area $A_h$ of the bow shock at the end of the jet,
and the ram pressure of the ambient medium. This yields (eqn. 1 of
BC)

\begin{equation}
v_h \sim \Bigl ( { Lj \over \rho _a v_j^3 A_h } \Bigr ) ^{1/2} v_j \> .
\end{equation}

Notice that $A_h$ can be much larger than the cross-section of the jet
itself (as in the `dentist drill' scenario (Scheuer 1982)). BC
inferred a value of $A_h \sim 30$ kpc $^2$ for Cygnus A from
observations. In their numerical simulation, Cioffi and Blondin (92) 
found that $A_h$ increases with time (roughly, $A_h \propto
t^ {0.4 \pm 0.1}$ depending on the value of $\eta$), and therefore,
$v_h$ decreases with time, albeit slowly. In the light of its
slow variation in time, and the uncertainties in other variables
involved in the problem, we will use a constant $A_h$ below.

The cocoon pressure $p_c$ is obtained simply from the total energy
deposited by the jet inside the volume of the cocoon $V_c = \epsilon_V
(2 \pi r_c^2) l_h$, where $r_c$ is the half-width of the cocoon
at the center, $l_h = \int v_h dt$ is the length of the jet head and
$\epsilon _V$ is a volume factor depending on the shape of the
cocoon (see Fig. 1). For a cylindrical cocoon, $\epsilon _V =1$ and 
for a biconical
shape, it is $\sim 1/3$ (Loken \etal 1992). One thus obtains
\begin{equation}
p_c \approx  { (\gamma -1) L_j t \over V_c } \approx \rho _a 
\Bigl ({ dr_c \over dt } \Bigr )^2 \> ,
\end{equation}
where the second equality comes from balancing the cocoon pressure
with the ram pressure of the ambient medium. $\gamma$ is the adiabatic
index of the material in the cocoon. This readily yields
upon integration an expression for the size of the cocoon, as
\begin{equation}
r_c^2 \approx  \Bigl ( {6 (\gamma -1) \over \pi} \Bigr )^ {1/2} 
\Bigl ( { L_j v_j A_h \over \rho _a } \Bigr 
)^ {1/4} \Bigl ({\epsilon _V \over 1/3} \Bigr ) ^{-1/2} t \> .
\end{equation}
Eqn $(1)$ and $(2)$ can then be combined to give the cocoon pressure, as
\begin{equation}
p_c \approx \Bigl ( {9 (\gamma -1) \over 24 \pi } \Bigr )^{1/2} 
\Bigl ( {L_j \rho_a \over v_h } \Bigr )^{1/2} \Bigl ({\epsilon _V \over 
1/3}\Bigr )^{-1/2} t^{-1} \>.
\end{equation}

Cioffi and Blondin (1992) argued on the basis of their simulation
that initially the cocoon energy is not totally thermalized and it
is about equally divided between thermal and kinetic energy. However,
at later times, the fraction of the thermalized energy approaches 
unity. 

The last equation (eqn ($4$)) allows us to estimate the time $t_{eq}$ for
the cocoon to reach pressure equilibrium with the ambient medium,
after which the cocoon boundary expands at the sound speed of
the ambient medium. For powerful radio galaxies, with length-scales
of the order of Mpc, the ambient medium is the IGM. We can write
the equilibrium time scale as,
\begin{eqnarray}
t_{eq}&\approx& {\mu m_p \over k T_a} \Bigl ( {L_j v_j A_h \over \rho_a
} \bigr ) ^{1/4} \Bigl ( {9 (\gamma -1) \over 24 \pi} \Bigr ) ^{1/2}
\Bigl ({\epsilon _V \over 1/3}\Bigr )^{-1/2} \nonumber\\
&\sim& 1.34 \times 10^{10} (T_{a,6})^{-1} \Omega_{IGM}^{-1/4} 
\> h_{50}^{1/2} \> L_{j,45}^{1/4}\> \beta_j^{1/4} A_{h,30}^{1/4}
\Bigl ({\epsilon _V \over 1/3}\Bigr )^{-1/2} \> {\rm yr}.
\end{eqnarray}
Here, $\mu \sim 0.6$ is the mean molecular weight, $T_a= 10^6 T_{a,6}$ K
is the temperature of the IGM gas, and $k$ is the Boltzmann's constant.
We have expressed $L_j$ and $A_h$ in the units of $10^{45}$ erg 
s$^{-1}$ and $30$ kpc$^2$ respectively, in accordance with BC. 
Whereas the earlier studies (BC; Cioffi and Blondin 1992, and Loken 
\etal 1992) used a value of $T_a \sim 10^8$ K, motivated by hot IGM models
for the X-ray background, thus obtaining 
$t_{eq} \sim 10^8$ yr, a much lower temperature of the IGM seems more
reasonable. The constraint from the COBE limit on the Compton
$y$ parameter ($y < 2.5 \times 10^{-5}$, Mather \etal 1993) and the
anisotropy measurements of the microwave background radiation in 
arcminute scales (${dT \over T} < 10^{-5}$, Subrahmanyan \etal 1993) pose 
severe problems with hot IGM models for the X-ray background (e.g., 
Loeb 1991). Moreover, success of the discrete source models in explaining 
the X-ray background (Zdziarski, \. Zycki, Krolik 1993) has 
diminished the motivation for such a hot IGM. Instead,
models of a photoionized IGM favour temperatures as low as $\sim
{\rm few} \times 10^4$ K (e.g., Miralda-Escud\`e and Ostriker 1990).
Therefore, eqn ($5$) suggests a equilibrium time scale
that can be even larger than the Hubble time ($\sim 1.3 \times 10^{10}
h_{50}^{-1}$ yr, for a $\Omega=1$ universe). The cocoons then
stay overpressured with respect to the IGM for all relevant
times in the evolution of the giant radio sources, unless $L_j, A_h$ 
have values much different than those used above.

We now turn our
attention to the determination of the ambient density of radio sources,
in particular, the IGM density from the
observations of giant radio sources.

\section{Intergalactic Matter}

The pressure jump across the cocoon boundary can be calculated
if the bow shock surrounding the supersonic jet is approximated
as a steady, oblique shock (Loken \etal 1992). The gas in the cocoon
is separated from the shocked ambient gas with pressure $p_{sh,a}$
just inside the bow shock (i.e., downstream)
by a contact discontinuity (i.e., $p_c \sim p_{sh,a}$). The pressure 
of the ambient medium $p_a$ and the cocoon pressure $p_c (=p_{sh,a})
$ are then related as (e.g., Landau and Lifshitz 1959),
\begin{equation}
{p_c \over p_a} = {5 M_h^2 \sin ^2 \phi -1 \over 4} \>,
\end{equation}
where $M_h $ is the Mach number of the head of the jet with respect
to the sound speed of the ambient medium ($c_{sa}$). In other words,
$M_h= v_h / c_{sa}$. $\phi$ is the angle that the bow shock makes
with respect to the jet axis and $\gamma=5/3$ is assumed for an
ideal gas in the above equation.

For a large pressure jump ($p_c / p_a \gg 1$), eqn ($6$) can be
rewritten in terms of the ambient density $\rho _a$,
\begin{equation}
\rho _a ={4 p_a \over 3 v_h^2 \sin ^2 \phi} \> .
\end{equation}
Furthermore, since $v_h=l_h/t$ (for a constant $v_h$), this
can also be written in terms of the age of the system $t$ and
the length of the jet $l_h$.  Note that, eqn ($7$) is essentially
eqn ($2$) in disguise: the transverse growth rate of the cocoon 
$({ dr_c \over dt })$ is replaced by a term proportional to
$(v_h \sin \phi)$, since the former is hard to estimate.

\medskip
\noindent
(i) Radio galaxies inside clusters: \medskip

Before going over to the case of giant radio galaxies embedded in the
IGM, let us use eqn $(7)$ to estimate the ambient density
for Cygnus A, which is inside a cluster, and whose ambient
density has been measured by X-ray observations. This will
be a consistency check for our method.
For Cygnus A, eqn ($7$) gives the ambient particle
density $n_a (= {\rho _a \over \mu m_p})$ as,
\begin{equation}
n_a \approx 2 \times 10^{-3} \Bigl ({p_c \over 8 \times 10^{-11}
}\Bigr ) \Bigl ({t \over 3 \times 10^7 yr} \Bigr)^2 \Bigl({l_h 
\over 60 \> kpc} \Bigr)^{-2} \Bigl ({\sin (20 ^\circ) \over 
\sin \phi }\Bigr)^2 \> .
\end{equation}
From X-ray observations, one estimates the ambient pressure
$p_a \sim 8 \times
10^{-11}$ dyne cm$^{-2}$ and an ambient electron density of $ \sim
6 \times 10^{-3}$ cm$^{-3}$ (Arnaud \etal 1987). 
A minimum age of $ \sim 3 \times 10^7
L_{j,45}^{-1}$ yr is estimated from the minimum total energy ($\sim
10^{60}$ ergs) in the radio lobes (BC). The pressure jump is,
therefore, of the order of $\sim 3$ if the angle of the bowshock
is $\sim 20^\circ$, for which the aspect ratio of the cocoon, i.e. the
ratio of the width to the length, is $\sim 0.4$.

\medskip
\noindent
(ii) Giant radio galaxies in the IGM: \medskip

Clearly, the angle $\phi$ is a difficult parameter to measure from
the radio observations of the lobes. However, for $\phi \la 30^\circ$
$\sin \phi \sim \tan \phi =r_c /l_h$, the difference being of the
order of unity. We can then use eqn ($3$) to eliminate
$\phi$ and rewrite the expression
for the ambient density. For giant radio galaxies, the ambient
density correspond to that of the IGM. We can express the
IGM density as,
\begin{equation}
\Omega_{IGM} \approx 0.01 h_{50}^{-2} (1+z)^{-3} \> l_{h, Mpc}^{4/3}\>
\Bigl ({p_c 
\over 10^{-14} \> {\rm d \> cm}^{-2}
} \Bigr )^{4/3} \Bigl 
({v_h \over 0.05 c}\Bigr)^{-4/3}
L_{j,45} ^{-1/3} \beta _j^{-1/3} A_{h,30}^{-1/3} \epsilon _{V ,1/3}
^{2/3} \>.
\end{equation}
We have written the length of the jet $l_h$ in the units of $1$
Mpc, $v_h$ in the units of $0.05c$ where $c$ is the velocity of light,
and the pressure $p_c$ in cgs units. From the statistics of radio sources,
Blandford and Rees (1984) estimated the limits of $v_h$ as being
$0.03 c < v_h < 0.1 c$. Therefore, an average value of $0.05c$ is 
reasonable. (Note that $v_h$ is not a constant for all galaxies,
as evident from eqn (1), and it increases with $L_j$. We have used
an average value in eqn (9) only to give a feeling for the numbers.
Also, note that there is no real evidence that such a value of $v_h$
is also reasonable for giant sources; but here we assume that it is.)
The pressure in the radio lobes of giant, with Mpc length
scales, radio sources at low redshifts ($z \la 0.1$) has been
estimated from minimum energy arguments, as being of the order of $10^{-14}$
dyne cm$^{-2}$ (Wagget, Warner and Baldwin 1977 obtained pressures
$\sim 5 \times 10^{-14}$ dyne cm$^{-2}$; but see Subrahmanyan and
Saripalli 1993). Recently, a giant source 8C 0821+695 at $z=0.538$ 
has been discovered with $l_h \sim 1.5 h_{50}^{-1}$ Mpc and
$p_c \sim 8 \times 10^{-14}$ dyne cm$^{-2}$ (Lacy \etal 1993).

An estimate of the jet luminosity ($L_j$) can be obtained from the 
product of the pressure in the hotspot, its apparent area and $v_j/2$
(e.g., Begelman, Blandford and Rees 1974).

It is worth while to note here that the estimates of the lobe pressure $p_c$
(from minimum energy argument) scale as $h_{50}^{4/7}$ and $l_h$ scales 
as $h_{50}^{-1}$. The overall dependency of $\Omega_{IGM}$ estimated from 
eqn ($9$) on the Hubble constant is then given by $h_{50}^{-18/7}$.

Requiring a overpressured cocoon, i.e. demanding $p_c \ga p_a$,
the lobe pressure observations can give an upper limit to the IGM pressure.
Using the above estimate of the IGM density, one can thus put an upper
limit to the temperature of the IGM gas, $T_{IGM}$. For Mpc long 
radio sources, with lobe pressures $\sim 10^{-14}$ dyne cm$^{-2}$ at the
present epoch, an upper limit of $T_{IGM} \la 10^7$ K is then implied.

To summarise this section: we have essentially used the transverse growth
rate of the cocoons of radio sources to estimate the ambient density (eqns
($2, 7, 9$)), since the expansion of the cocoon in that direction is
determined by the balance between the cocoon pressure ($p_c$) and the
ram pressure of the ambient medium. We have used the shock jump
conditions to relate the jet head velocity ($v_h$) and the aspect ratio
of the cocoon, to the transverse growth rate, leading to eqns ($7$) and 
($9$). For giant radio sources, this gives an estimate of the IGM
density as a few percent of the closure density of the universe.

We will postpone a discussion on the cosmological implications of this 
estimate of the IGM density till $\S 5$, and calculate the radio 
luminosity of the overpressured cocoons in the next section. We will
also discuss in $\S 5$ the uncertainties in the estimates of the
lobe pressure that we have used above.


\section{Radio Luminosity}

The radio power of lobes which are expanding with an equal pressure
($p_a \sim p_c$) or expanding freely, with a subsonic velocity, 
has been calculated by various previous authors (e.g., Eilek and Shore 
1989). We will calculate the radio luminosity in the case of the 
overpressured cocoons discussed above, mainly the way it scales with 
various parameters. We will follow the method outlined in Daly (1994).

The radio luminosity $L_r$ of the cocoon at frequency $\nu$ is 
approximately $L_r( \nu ) \approx \nu P_r (\nu )$, where $P_r$ is
the radio power at $\nu$. If the rate of synchrotron emission
per relativistic electron is $dE/dt$ at the frequency $\nu$ and
the number of relativistic electrons with energy $E$ is $N(E)$, then
the radio luminosity at the frequency $\nu$ can be approximately
written as $L_r (\nu ) \approx N(E) dE/dt$. The energy $E= \gamma m_e 
c^2$, where $\gamma$ is the Lorentz factor. For a spectral
index $\alpha$ at the radio frequencies (i.e., the radio luminosity
is proportional to $\nu^{- \alpha }$), $N(E) = N_{tot} (\gamma /
\gamma _l)^{-2 \alpha }$, where $\gamma _l$ denotes the lower cutoff
in energy. $N_{tot}$ is the total number of relativistic electrons
in the cocoon, and in the limiting case that the total cocoon
energy is in these electrons, can be written as $N_{tot} \approx (L_j 
t / \gamma _l m_e c^2)$.

The synchrotron emission rate $dE/dt \approx 1.6 \times 10^{-15}
\gamma ^2 B^2$, where the relativistic electron spirals around 
a magnetic field with $B_{\perp}$ as the component perpendicular to
its velocity. Assuming equipartition between magnetic and gas
energy, the magnetic field in the cocoon can be written (from
eqn ($3$)) as $B \sim \Bigl ( { 16 \pi \rho _a L_j \over v_h} 
\Bigr )^{1/4} t^{-1/2}$. This readily leads to the radio
power at the frequency $\nu$, given by
\begin{equation}
P_r(\nu ) \sim {5 \times 10^{-8} \over (4 \times 10^6)^{(1- \alpha)}
} (16 \pi)^{{1+ \alpha \over 4}} L_j^{{5+ \alpha \over 4}}
\rho _a^{{1+ \alpha \over 4}} v_h^{-{1+ \alpha \over 4}} \gamma _l^
{(2 \alpha -1)} \nu ^{- \alpha} t^{{1- \alpha \over 2}} \>.
\end{equation}
This predicts a small increase in the radio power 
with time for $\alpha \sim 1$ and with the ambient density $\rho _a$.
Radio lobes in higher density environments would therefore have
larger total luminosities. One should however remember that there
is no direct evidence for the equipartition of energy that we
have assumed (see below).

Note that $P_r$ above is the total
radio power of the cocoon. As emphasized by Loken \etal (1992),
the radio luminosity will not be uniform throughout the cocoon. If the
last reacceleration of the electrons happens at the Mach disk, the
backflow velocity ($\sim$ sound speed in the cocoon, $c_s$) 
and the magnetic field $B$
would limit the size of the lobe observed at a certain frequency $\nu$.
The size of the lobe will be $\sim c_s \tau _s$, where $\tau _s$ is the
synchrotron life time of relativistic electrons with energies 
corresponding to the observed frequency, in the presence of
magnetic field $B$. 

However, it is possible that one will observe more extensive cocoons
if reacceleration also happens elsewhere. It is also possible that if
one uses beams that are sensitive enough to detect large structures,
one could observe limb brightening at the cocoon boundary. This would
provide support of the BC model for extensive cocoons.

It is interesting to note here that, at a given redshift, or 
in other words, at a constant $\rho _a$, $v_h \propto Lj^{1/2}$ (eqn 
($1$)), which means that for a constant average $v_h$, the length of the
radio source $l_h \propto v_h \propto L_j^{1/2} \propto P_r(\nu)
^{{2 \over 5 + \alpha}}$. For $\alpha \sim 1$, this recovers
the correlation $l_h \propto P_r ^{0.3}$ that some observers
claim to have found (e.g., Oort \etal 1987). Other
studies have also attempted to explain this correlation, although
with specific models of structure of the ambient matter (Gopal-Krishna
and Witta 1991).


\section{Discussion}
While using the lobe pressure $p_c$ in determining $\Omega _{IGM}$
 as in $\S 3$, the
assumptions behind the estimates of $p_c$ must be clearly borne in mind.
Lobe pressure is usually calculated from the minimum
energy arguments, which gives roughly the equipartition values.
One of course does not have any direct evidence that equipartition holds 
in all cases, although sources embedded in clusters seem to yield
pressure estimates that tally with X-ray
observations. Also, the efficiency of energy conversion into
relativistic electrons and magnetic field is not known. All
these contribute to the uncertainty in the lobe pressure
estimates. 

It is also important to note that in reality most of the powerful 
radio galaxies are asymmetric and the diagram in Fig. 1
should be treated as being only schematic in nature. McCarthy, 
van Breugel and
Kapahi (1991) have examined nonthermal radio emission from the
plasma and line emissions from the thermal gas in a few of these
asymmetric sources and suggested that the asymmetry is caused
by density inhomogeneity.

Our estimate of the IGM density ($\S 3$) has important
cosmological implications. First we note that
a direct estimation of the density of IGM $\Omega_{IGM}$ has not
been possible in the past and only indirect arguments has been put 
forward to  place various limits (e.g., Barcons, Fabian and Rees (1991)). 
Primordial nucleosynthesis model limit the fraction of 
the baryonic density of the universe to $\Omega_b \sim (0.05 \pm
0.01) h_{50}^{-2}$ (Walker \etal 1991). The luminous matter in 
galaxies at $z=0$ contribute about $\Omega_{gal} \sim 0.007$,
a similar amount in intracluster gas and some galactic or cluster
baryonic dark matter may be present( Barcons, Fabian and Rees (1991)).
They argued that a value of $\Omega_{IGM} \ga 0.01$ is 
therefore reasonable. We note that our estimate of $\Omega_{IGM}$ 
from eqn ($9$) is consistent with
these arguments. 

Clearly, a homogeneous IGM is an ideal case and in reality the 
universe is probably clumpy. The size and magnitude of the clumpiness
are still beyond our knowledge, though the clumping scale is limited
by the upper limits of the anisotropy of the microwave background
radiation, especially in the arcminute scale (Barcons, Fabian and
Rees 1991). It is interesting to note
that the estimates of lobe pressure in Mpc length radio sources
do not vary much. Whether this indicates a selection effect
or not, needs to be confirmed by looking for giant sources with low
surface brightnesses (Lacy \etal 1993). In any case, 
our estimates refer to the density 
in the vicinity of the radio source only and one ought to be
cautious in drawing conclusions about the global homogeneity
of the universe from a handful of giant radio sources.

If the IGM is rather uniform, then,
one can constrain the temperature
of the IGM gas from the limits on Compton $y$ parameter from COBE
along with the knowledge of $\Omega_{IGM}$
(as in Barcons, Fabian and Rees 1991, though the limit they used
is now outdated by almost two orders of magnitude). However, this
depends on the temperature and ionization history of the IGM,
especially at high redshifts. Their argument against a hot
IGM model for the production of the X-ray background radiation
would however be strengthened by our estimate of $\Omega_{IGM}$ by
narrowing the possible range of its value.


\section{Conclusion}
We have outlined a method to estimate the density of the
intergalactic medium from observations of giant radio
sources, in the framework of the model of overpressured cocoons surrounding 
these sources. We estimated an IGM density that is a few percent of the
closure density of the universe. We also discussed the radio luminosity of 
the cocoons and found the model consistent with the observed
correlation $l \propto P_r ^{0.3}$.

\bigskip

\centerline{\bf Acknowledgements}

I am indebted to Drs. Mitchell Begelman, Peter Biermann, Ruth Daly, 
Peter Scheuer and an anonymous referee
for their comments on the manuscript. I have benefitted much
from discussions with Heino Falcke, Lakshmi
Saripalli and Ravi Subrahmanyan.
This work was supported in part by the Max Planck Society and
a fellowship from IUCAA.

\newpage
\noindent
{\bf Figure Caption}

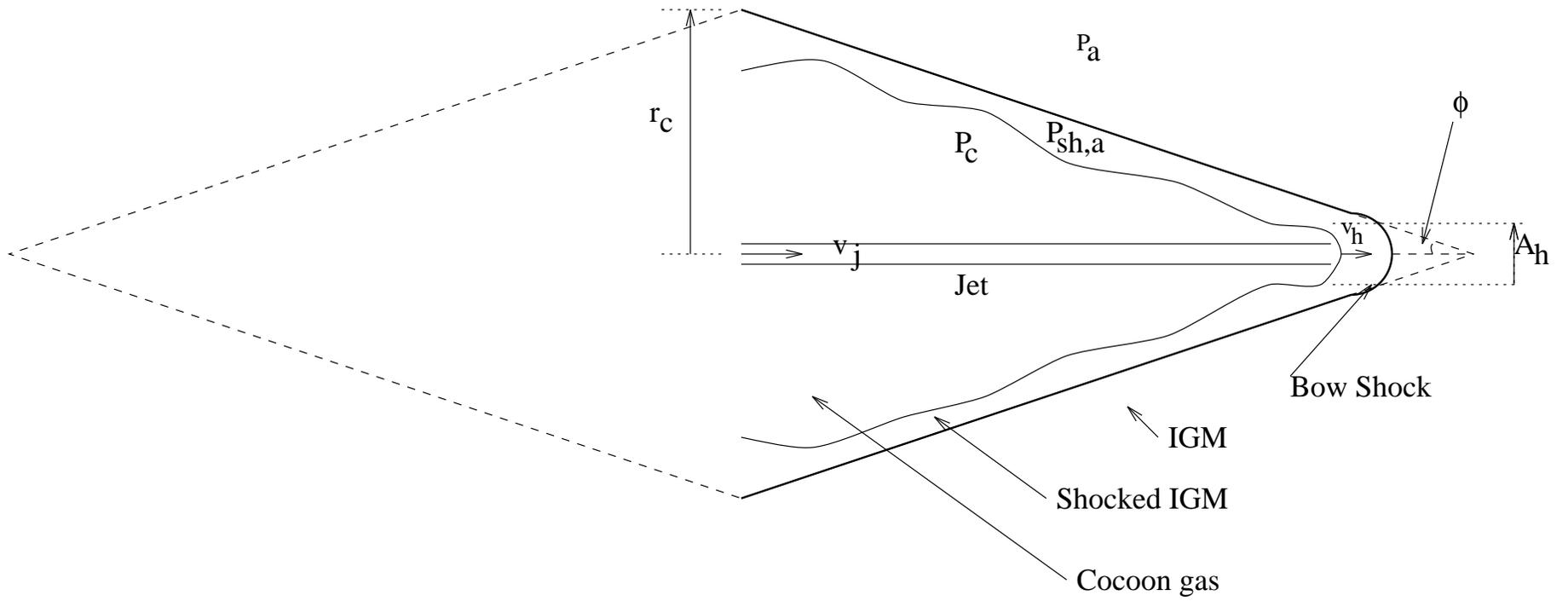
Figure 1 -- A schematic diagram of the cocoon surrounding a supersonic
jet (see also the fig. 1 in BC and Loken \etal 1993). The jet
moves with velocity $v_j$ and the head of the jet advances
in the ambient medium (here the IGM) with velocity $v_h$ and
a with a cross sectional area $A_h$. The bow shock makes an
angle $\phi$ with the jet axis and separates the ambient
medium (with pressure $P_a$) from the shocked ambient medium
(with pressure $P_{sh,a}$). There is a contact discontinuity
between the shocked ambient medium and the cocoon gas (with
pressure $P_c$). The width of the cocoon at the centre is
$r_c$.

\end{document}